\documentclass[%
 reprint,
superscriptaddress,
nofootinbib,
 amsmath,amssymb,
 aps,
 showkeys,
]{revtex4-1}
\usepackage{graphicx}
\usepackage{dcolumn}
\usepackage{bm}
\usepackage{hyperref}
\usepackage[bottom]{footmisc}
\usepackage{footnote}
\usepackage{xcolor}
\usepackage[mathlines]{lineno}
\usepackage{fontenc}
\usepackage{graphicx}
\usepackage{dcolumn}
\usepackage{bm}
\usepackage{booktabs}
\usepackage{array,multirow}
\usepackage{amsmath}
\usepackage{amsfonts}
\usepackage{amssymb}
\usepackage{mathrsfs}
\usepackage{enumerate}
\usepackage{fancyhdr}
\usepackage{xcolor}
\usepackage{graphicx}
\usepackage{listings}
\usepackage{float}
\usepackage{verbatim}
\usepackage{braket}
\usepackage{centernot}
\usepackage{setspace}
\usepackage{endnotes}
\usepackage[normalem]{ulem}
\usepackage{wrapfig}
\usepackage{multirow}
\usepackage{comment}
\usepackage{kantlipsum}
\allowdisplaybreaks
\usepackage{lipsum}
\usepackage[english]{babel}
\hypersetup{breaklinks = true, colorlinks = true, citecolor = blue, linkcolor = blue, urlcolor = blue}
\usepackage[mathlines]{lineno}

\newcommand{\be}{\begin{equation}}
\newcommand{\ee}{\end{equation}}
\newcommand{\bea}{\begin{eqnarray}}
\newcommand{\eea}{\end{eqnarray}}

\begin{document}

\title{Krylov complexity and gluon cascades in the high energy limit}

\author{Pawel Caputa}
\email{pawel.caputa@fuw.edu.pl}
\affiliation{Faculty of Physics, University of Warsaw, Pasteura 5, 02-093 Warsaw, Poland}
\affiliation{Yukawa Institute for Theoretical Physics, Kyoto University, Kitashirakawa Oiwakecho, Sakyo-ku, Kyoto 606-8502, Japan}

\author{Krzysztof~Kutak}
\email{krzysztof.kutak@ifj.edu.pl}
\affiliation{Institute of Nuclear Physics, Polish Academy of Sciences, ul. Radzikowskiego 152, 31-342,
Krak\'ow, Poland}


\begin{abstract} 
We point out an interesting connection between the mathematical framework of the Krylov basis, which is used to quantify quantum complexity, and the entanglement entropy in high-energy QCD. In particular, we observe that the cascade equation of the dipole model is equivalent to the $SL(2,R)$ Schrodinger equation in the Krylov basis. Consequently, the Krylov complexity corresponds to the average distribution of partons and the Krylov entropy is the counterpart of the entanglement entropy computations of \cite{Kharzeev:2017qzs}. Our work not only brings new tools for exploring quantum information and complexity in QCD, but also gives hope for experimental tests of some of the recent, physical probes of quantum complexity.  

\end{abstract}

\keywords{Complexity, Entanglement entropy, QCD}
\maketitle
 
\begin{figure}[b!]
    \centering
    \includegraphics[width=0.39\textwidth]{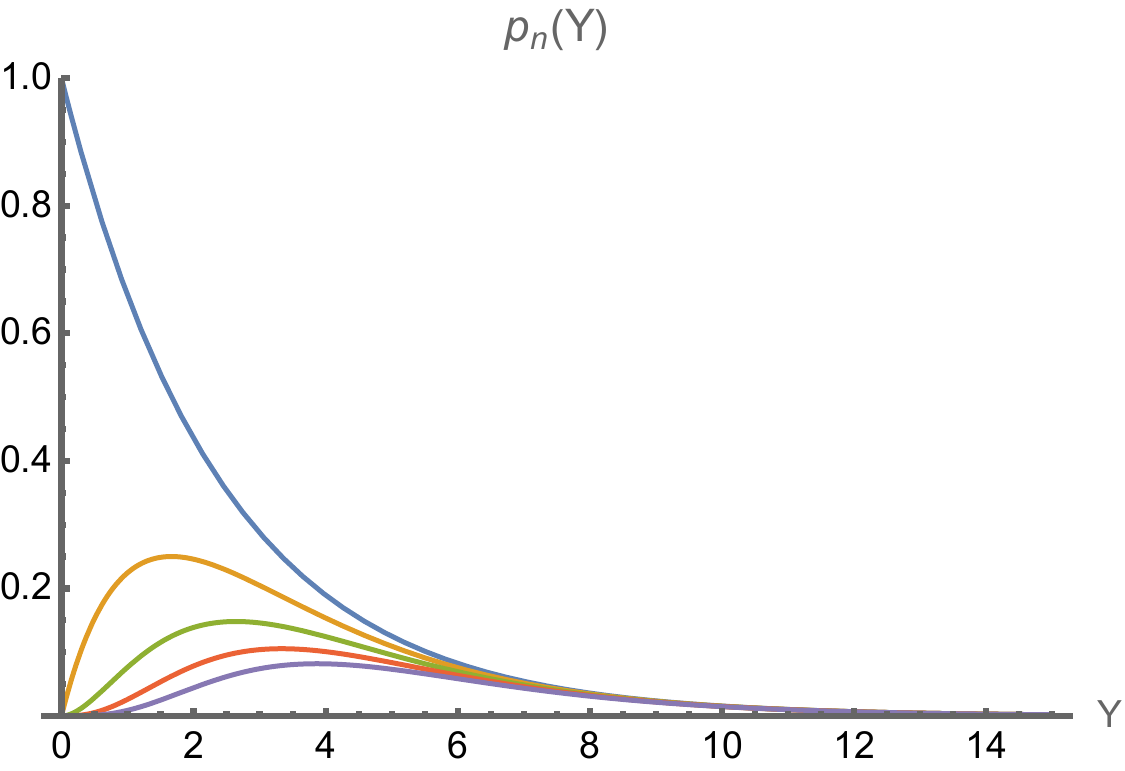}
    \caption{Solutions of the equation (\eqref{eq:cascade1}) for $p_n$ with $n$ from 1 (the top blue line) to 5 (the purple line at the bottom) for $\alpha_s=0.15$. }
    \label{fig:results}
\end{figure}

\paragraph*{\textbf{Introduction.}}
Recently, quantum information (QI), with concepts such as entanglement or complexity, has provided a new perspective on high-energy physics. This is particularly fruitful for understanding quantum black holes in the AdS/CFT correspondence \cite{Maldacena:1997re} (see e.g.\,\cite{Almheiri:2020cfm}). Moreover, precise measures of quantum complexity are also believed to be crucial for understanding black hole interiors \cite{Susskind:2014rva,Brown:2015bva}. These exciting developments have motivated a new paradigm of exploring quantum systems from the angle of entanglement and complexity.\\
In fact, in a series of papers \cite{Kutak:2011rb,Peschanski:2012cw,Stoffers:2012mn,Armesto:2019mna,Kharzeev:2017qzs,Kovner:2018rbf,Kovner:2022jqn,Liu:2022ohy,Liu:2022bru,Dumitru:2022tud,Kharzeev:2021nzh,Liu:2022hto,Liu:2023zno,Liu:2022qqf,Liu:2023eve,Dumitru:2023fih,Chachamis:2023omp,Kutak:2023cwg} the authors argued that understanding entanglement in QCD may provide new insights into the gluon-dominated structure and properties of high-energy hadrons. 
This was supported by the computation of the entanglement entropy in the electron-proton Deep Inelastic Scattering (DIS) process \cite{Kharzeev:2017qzs}, which is based on the following logic. The proton is an eigenstate of the QCD Hamiltonian (a pure state), however, a measurement of its partonic content at some resolution scale introduces a bipartition between resolved and unresolved parts that are entangled in an intricate way. This entanglement can be quantified by the von Neumann entropy 
\begin{equation}
S_A=-\text{Tr}(\rho_{A}\log\rho_{A})=-\sum_n p_n\ln p_n,\label{SavN}
\end{equation}
where $\rho_A=Tr_B(\rho_{AB})$ is the reduced density matrix of the resolved part $A$ ($B$ is its complement). The eigenvalues $p_n$ are interpreted as probabilities of a state with definite number ($n$) of partons/dipoles. \\
More importantly, \cite{Kharzeev:2017qzs} conjectured that this entanglement entropy is directly related to the hadronic entropy calculated with the Gibbs formula after counting the hadrons measured in the collider experiment. The motivation for this conjecture comes from the parton-hadron duality \cite{Dokshitzer:1987nm}, where hadrons can be directly linked to partons, which fragment and hadronize.  
Following this proposal, the measurement of hadronic entropy in DIS has been performed \cite{H1:2020zpd} and seems to be consistent with the theoretical prediction \cite{Hentschinski:2021aux,Hentschinski:2022rsa,Kou:2022dkw}. Further evidence for the general picture of the proton as the maximally entangled state comes from studies of proton-proton collisions \cite{Tu:2019ouv} and diffractive DIS \cite{Hentschinski:2023izh}, where the hadronic entropy has been successfully described.\\ 
In this work, we shed new light on this story by pointing out its mathematical equivalence to the Krylov space approach to quantum complexity \cite{Parker:2018yvk,Caputa:2021sib,Balasubramanian:2022tpr}. Intuitively, it has been observed that the dynamics of complexity and growth of quantum operators are well captured by epidemic/cascade models \cite{Susskind:2014jwa,Qi:2018bje}, and the parton evolution equations that we will consider are examples thereof. Having established the precise connection, we will use it to bring new tools such as the capacity of entanglement, K-variance or continuum limits to the QCD discussions and to probe the saturation effects.  
\paragraph*{\textbf{Dipole cascade model.}}
In this section, we review the main findings of \cite{Kharzeev:2017qzs}. In QCD at high energies, the convenient degrees of freedom of the proton are color dipoles and quadrupoles \cite{Mueller:1985wy,Kovchegov:2012mbw,Dominguez:2012ad}. 
The latter can be expressed, in the approximation of a large number of colors, as dipoles that represent a pair of high-energy quarks or gluons. The usefulness of this representation derives from the fact that, at high energies, the dipole size is fixed during the interaction time and the $S$-matrix of the interaction is diagonal with respect to the transverse dipole size \cite{Kovchegov:2012mbw}.
 In \cite{Kharzeev:2017qzs,Liu:2022hto,Liu:2022bru}, the authors considered two scenarios: the 3+1 D Mueller's picture (see \cite{Kovchegov:2012mbw} for a review) \cite{Mueller:1994gb} as well as its 1+1 D reduction. Here we focus on the 1+1 D case, which is exactly solvable and, in the low $x$ regime (explained below), approximates the Balitski, Fadin,
Kuraev, Lipatov (BFKL) growth of the gluon density
\cite{Balitsky:1978ic,Kuraev:1977fs,Kwiecinski:1997ee,Hentschinski:2013id}.\\
The probability $p_n(Y)$ of finding a state with $n$ dipoles at rapidity $Y$ (playing the role of the evolution time) obeys the equation \cite{Mueller:1994gb,Levin:2003nc}
\be
\label{eq:cascade1}
\partial_Y p_n(Y)=-\lambda\,n\,p_n(Y)+\lambda\,(n-1)\,p_{n-1}(Y),
\ee
where the parameter $\lambda$ characterizes the speed of the evolution which in the  BFKL equation is a prediction and can be obtained from the numerical solution \cite{Askew:1993jk}.\\
 The first term in \eqref{eq:cascade1} describes the decrease in the probability due to the splitting into (n+1) dipoles, while the second one describes the increase due to the splitting of (n-1) dipoles into n dipoles.
The solutions of \eqref{eq:cascade1} are given by
\be
p_n(Y)=e^{-\lambda Y}\left(1-e^{-\lambda Y}\right)^{n-1},\label{SolMul}
\ee
and we plot some of them in fig.\,\ref{fig:results}. 
At large rapidities, the probabilities tend to the same value, which is interpreted as a state of maximum entanglement \cite{Kharzeev:2017qzs,Hentschinski:2023izh}. 
We used  $\lambda=0.4$, which, via formula $\lambda=4\bar\alpha_s \ln 2$, as obtained from BFKL in diffusion approximation \cite{Askew:1993jk}. The value we use follows from assumption that QCD coupling constant is fixed at value $\alpha_s=0.15$, $\bar\alpha_s=N_c\alpha_s/\pi$, where $N_c=3$ is a number of colors. The larger value of the coupling constant leads to faster evolution, but  the features of the solutions are essentially unchanged. The entropy \eqref{SavN} associated with the system of dipoles is given by \cite{Kharzeev:2017qzs}
\begin{equation}
    S=
\lambda Y+(1-e^{\lambda Y}) \cdot \log\left(1 - e^{-\lambda Y}\right),\label{EEQCD}
\end{equation}
while the average number of dipoles grows exponentially with rapidity
\be 
\langle n\rangle=\sum^\infty_{n=1} n p_n(Y)= e^{\lambda Y}.
\label{eq:meangluons}
\ee
The equation above,  where  $Y=\ln 1/x$ with Bjorken $x$ $-$ longitudinal momentum fraction of the proton as carried by the parton, is a model for the BFKL momentum density of partons and it features the growth of the number of partons (mainly gluons) with increasing rapidity.
Taking the high energy limit of eq. \eqref{EEQCD} and combining it with eq. \eqref{eq:meangluons} one gets the relation between entanglement and parton distribution ($g(x)$ - denoting the number of gluons at a given x) \cite{Kharzeev:2017qzs} 
\begin{equation}
    S\simeq\lambda Y=\ln \langle n\rangle\simeq \ln xg(x).
\end{equation}
In the above, the parton momentum density has been represented by gluons. 
The generalization  accounting for quarks and the charge factor can be found in \cite{Hentschinski:2021aux}.

Let us now switch for a moment to a seemingly distant and unrelated corner of the high-energy theory, the Krylov complexity.
\paragraph*{\textbf{Krylov Complexity.}}
Complexity is usually quantified using a circuit model, where the circuit that performs a given task is constructed from a universal set of gates. The complexity of the task is defined as the minimum number of gates used to construct the circuit that accomplishes it \cite{shannon1949synthesis}. Despite this simple intuition, it is very difficult to precisely quantify complexity in quantum systems, especially in quantum field theories. \\
Nevertheless, a new physical measure of  complexity of time evolution, defined as a spread in Hilbert space, has recently been developed at the interplay between many-body and high-energy physics. It started from the observation that epidemic models capture very well the growth of operators and the evolution of complexity in holography \cite{Susskind:2014jwa,Qi:2018bje}. This was sharpened using the Krylov basis \cite{Parker:2018yvk} and extended to a definition of state complexity in \cite{Balasubramanian:2022tpr} (see also e.g. \cite{Rabinovici:2020ryf,Dymarsky:2021bjq,Caputa:2021ori,Caputa:2022eye,Rabinovici:2023yex,Jian:2020qpp,vonKeyserlingk:2017dyr}). 
The key point is to 
expand the state or the operator in the minimal basis that supports its unitary evolution (the so-called Krylov basis). To be concrete, we examine it for the state $\ket{\Psi(t)}$ that solves the Schrodinger equation with a general time-independent Hamiltonian $H$ 
\be
\ket{\Psi(t)}=e^{-iHt}\ket{\Psi_0}=\sum_{n=0}\psi_n(t)\ket{K_n}.\label{State1}
\ee
The Krylov basis $\ket{K_n}$ is constructed by the iterative Gram-Schmidt procedure (the Lanczos algorithm \cite{Lanczosbook}) applied to the Krylov subspace consisting of the initial state $\ket{\Psi_0}\equiv \ket{K_0}$ and all powers of the Hamiltonian acting on it $H^n\ket{\Psi_0}$. 
The Lanczos algorithm produces a basis in which the Hamiltonian is tri-diagonal
\be
H\ket{K_n}=a_n\ket{K_n}+b_n\ket{K_{n-1}}+b_{n+1}\ket{K_{n+1}},\label{KBdef}
\ee
with $a_n$ and $b_n$, called  Lanczos coefficients, obtained in the iterative procedure. Consequently, the amplitudes $\psi_n(t)$ in \eqref{State1} satisfy the Schrodinger equation
\be\label{SchrodingerEq}
i\partial_t\psi_n(t)=a_n\psi_n(t)+b_{n+1}\psi_{n+1}(t) + b_n\psi_{n-1}(t) \;.
\ee
Effectively, this approach maps the evolution to a one-dimensional chain with sites labeled by the index $n$ of the Krylov basis vectors, and encodes the dynamics into the probability distribution $p_n(t)\equiv |\psi_n(t)|^2$.
This also brings us to the key concept of the Krylov complexity (K-complexity for short), which is defined as the average position $n$ on the chain
\be
\mathcal{C}_K(t)=\langle n\rangle=\sum_n n\,p_n(t).
\label{eq:krylovC}
\ee
Of course, there is more information about the dynamics that is hidden in $p_n(t)$ than just in $\langle n\rangle$ and we often employ additional QI tools to probe it. The first one is the Shannon entropy measuring the information content of the distribution (dubbed the K-entropy \cite{Barbon:2019wsy}) 
\be
S_K(t)=-\sum_n p_n(t)\log p_n(t).\label{eq:KEntropy}
\ee
Moreover, with the K-variance we can quantify the fluctuations around the average position $\langle n\rangle$  \cite{Caputa:2021ori}
\be
\delta^2_K=\frac{\langle n^2\rangle-\langle n\rangle^2}{\langle n\rangle^2}.\label{KVariance}
\ee
In addition, from the m-th Renyi entropy
\be
S^{(m)}_K=\frac{1}{1-m}\log\left(\sum_n p^m_n(t)\right),
\ee
we can define the analog of the heat capacity, the Capacity of Entanglement \cite{Yao_2010,DeBoer:2018kvc,Kawabata:2021hac}
\be
C_E=\lim_{m\to1}m^2\partial^2_m[\left(1-m\right)S^{(m)}_K].\label{eq:KCapacity}
\ee
The last quantity that we will consider is the purity, defined as 
\be
\gamma_K=\sum_{n}p^2_n(t).\label{eq:Purity}
\ee
We will use these tools for dipole cascades after making a precise connection below.\\
Most importantly, equation \eqref{SchrodingerEq} admits a class of analytic solutions with $a_n=0$ and $b_n=\alpha n$ for some constant $\alpha$ (see Appendix A) given by
\be
\psi_n(t)=(-i)^n\frac{\tanh^n(\alpha t)}{\cosh(\alpha t)}.\label{SolSL2R}
\ee
They were first derived in \cite{Roberts:2018mnp,Parker:2018yvk} for the Sachdev-Ye-Kitaev (SYK) model \cite{Sachdev:1992fk,kitaev} and later explained using the dynamical SL(2,R) symmetry in \cite{Caputa:2021sib}. Interestingly, we can associate them with \eqref{State1} evolved by the boost-type Hamiltonian 
\be
\ket{\psi(t)}=e^{-i\alpha(L_1+L_{-1})t}\ket{0}\otimes\ket{0},
\ee
where $L_{\pm 1}$ are the SL(2,R) generators and $\psi_n(t)$ follow from the Baker–Campbell–Hausdorff relation for this Lie algebra. If we express them in the two-mode representation (see Appendix A), they are nothing else but the Schmidt coefficients in this two-mode decomposition.\\ 
The K-complexity for this solution grows exponentially with time
\be
\mathcal{C}_K(t)=\sum^\infty_{n=0} n|\psi_n(t)|^2=\sinh^2(\alpha t),
\ee
while the K-entropy \eqref{eq:KEntropy} (see also \cite{Patramanis:2021lkx}) 
\be
S_K(t)=2\log(\cosh(\alpha t))-2\log\left(\tanh(\alpha t)\right)\sinh^2(\alpha t),
\ee
increases linearly for late times, leading to the relation $ \mathcal{C}_K(t)\sim \exp(S_K(t))$.\\
We conclude this part with the observation  \cite{Parker:2018yvk} that from the exponential growth of $\mathcal{C}_K(t)$ at late times we can define a Lyapunov exponent $\lambda_K=2\alpha$. In fact, this time growth is correlated with the asymptotic growth of the Lanczos coefficients $b_n$ with $n$, which was conjectured to be at most linear 
\be
b_n\sim\alpha\, n+ \kappa,\label{bnbound}
\ee
where $\alpha$ and $\kappa$ are system-dependent constants. For example, in the SYK model this bound is saturated and the Krylov-Lyapunov exponent coincides with the maximum chaos bound \cite{Maldacena:2015waa} from the out-of-time-ordered correlators $\lambda_K=\lambda_{OTOC}=2\pi/\beta$, where $\beta$ is the inverse temperature $\beta=1/T$ of the system.
\paragraph*{\textbf{The Connection.}}
In this section, we present the main results of our work. First, we observe that \eqref{SchrodingerEq} implies an equation for the probabilities $p_n(t)$ (in complete analogy with the usual Schrodinger and continuity equations). More precisely, we can show (see appendix A) that $p_n(t)=|\psi_n(t)|^2$ associated with the solution \eqref{SolSL2R} satisfy the cascade equation
\be
\partial_Y p_n(Y)=\alpha n\, p_{n-1}(Y)-\alpha (n+1)\,p_{n}(Y),\label{EqProbY}
\ee
where we have introduced a ``rapidity" variable $Y$ related to the time $t$ via
\be
\tanh^2(\alpha t)=1-e^{-\alpha Y}.
\ee
This is almost the same as \eqref{eq:cascade1}, and the only difference is the range of the index $n$. Namely, in the QCD model $n$ is the number of dipoles $n\in(1,\infty)$, and in the Krylov basis we have $n\in(0,\infty)$. Indeed, by shifting the index $n\to n+1$, we verify that the SL(2,R) solution expressed in terms of $Y$ satisfies the cascade equation \eqref{eq:cascade1} after identifying $\alpha$ with $\lambda$. 
Consequently, the average dipole number $\langle n\rangle$ maps to the Krylov complexity (more precisely $\mathcal{C}_K+1$)
\footnote{In \cite{Liu:2022hto} the summation index $n$ is shifted so that in this case the relation to Krylov complexity does not need shift. The arbitrariness is related to choice of initial  condition and is not relevant at high energy limit.}. 
Given the above and using 
eq. \eqref{eq:meangluons}, we can write the relation between the K-complexity (in the cascade model) and the parton distribution
\be
 \mathcal{C}_K=xg(x).
\ee
We also note that the entanglement entropy \eqref{EEQCD} is the counterpart of the K-entropy.
These connections are one of our main results\footnote{We should also point out that \eqref{SolMul} appeared in the context of the operator growth in SYK  \cite{Roberts:2018mnp}, but $Y$ was replaced by $t$ and there was no relation with the Krylov approach.}.\\
Finally, we point out that the coefficients in the cascade equation, eq.\,\eqref{eq:cascade1}, satisfy the bound \eqref{bnbound} (choosing $\alpha=\lambda$ and $\kappa=0$). Whether this  reflects any "chaotic" or scrambling properties of QCD, i.e., $\lambda=4\bar\alpha_s \ln 2$ plays the role of the Lyapunov exponent, is beyond the scope of this work, but it would be fascinating to explore these questions, given the bulk of developments following \cite{Maldacena:2015waa}.\\
After explaining the relation between these two developments, in the next sections we will study a generalized dipole evolution equation that takes into account saturation effects and probe it with QI tools from the Krylov approach. Understanding the possible mechanisms behind both saturation of parton density and  saturation of complexity is very important and we will explore them in this tractable QCD scenario. 

\paragraph*{\textbf{Dipole cascade with recombinations.}}
\begin{figure}[t]
    \centering
    \parbox{0.39\textwidth}{\includegraphics[width=0.39\textwidth]{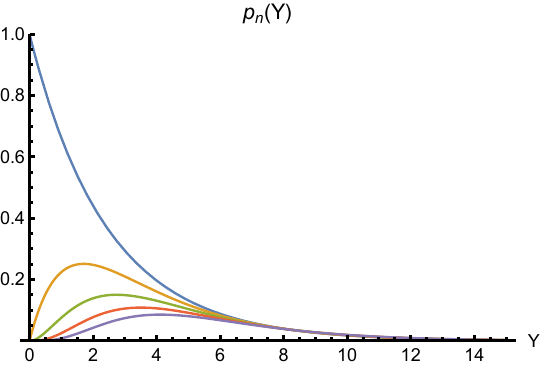}}$\,$ \parbox{0.39\textwidth}{\includegraphics[width=0.39\textwidth]{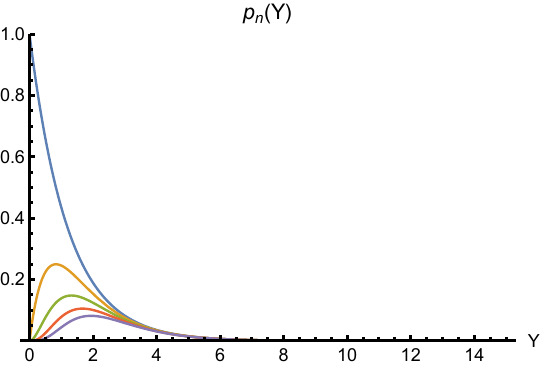}}
    \caption{ Solutions of the equation (\eqref{eq:cascade2}) for $p_n$ from 1 (the top blue line) to 5 (the bottom purple line). Upper plot:  $\alpha_s=0.15$. Lower plot: $\alpha_s=0.3$. }
    \label{fig:results2}
\end{figure}
In QCD at high energies, once the parton density is high enough, gluons are expected not only to split but also to recombine. The recombination changes the exponential growth of the number of gluons (with rapidity) and leads to saturation \cite{Gribov:1983ivg,Mueller:1985wy} (see also   \cite{Gelis:2010nm,Kovchegov:2012mbw} for a review). While there are hints of gluon saturation \cite{Kovchegov:2012mbw,Albacete:2014fwa},  strong evidence is still lacking and the search for saturation is the subject of forthcoming programs at both the Electron Ion Collider
\cite{AbdulKhalek:2021gbh} and the Large Hadron Collider \cite{Morreale:2021pnn,vanHameren:2023oiq}.
The evolution equations 
that account for gluon saturation are nonlinear generalizations \cite{Kovchegov:1999yj,Kovchegov:1999ua,Balitsky:1995ub,Iancu:2001ad,Jalilian-Marian:1997jhx,Jalilian-Marian:1997qno} of the BFKL equation. The 1+1 dipole model can also be generalized to account for recombination that leads to saturation. The generalized evolution equation reads \cite{Shoshi:2005pf,Bondarenko:2006rh}
\begin{align}
\label{eq:cascade2}
    \partial_Y p_n(Y) &= -\lambda n p_n(Y) + \lambda(n - 1)p_{n-1}(Y) \\\nonumber
                      &\quad + \beta n(n + 1)p_{n+1}(Y) - \beta n(n - 1)p_n(Y),
\end{align}
where $\beta=\lambda \alpha_s^2$. Clearly, the first two terms are the same as in \eqref{eq:cascade1}, while the last two account for dipole recombination. The proportionality of the recombination terms with $\alpha_s^2$ indicates that they are relevant at large values of rapidity, when the smallness of the coupling constant is compensated by the large number of dipoles.\\
This equation was solved analytically in the approximation of large $Y$ in \cite{Hagiwara:2017uaz}, but in order to test our new QI tools in this context, we simply solve it numerically with initial conditions $p_1(0)=1$ and $p_n(0)=0$ for $n>1$ (see also Appendix B for solutions in the continuum limit).
The resulting probabilities are plotted in fig.\,\ref{fig:results2} and the curves are similar to those in fig.\,\ref{fig:results}. Nevertheless, the behavior changes drastically as we increase the value of the coupling $\alpha_s$. In this regime, the term responsible for the dipole recombination starts to play an important role already at lower rapidities. It also leads to a scenario  where at asymptotic rapidities the probability with a large number of dipoles dominates and even increases with increasing rapidity.
\paragraph*{\textbf{QI tools for dipole cascades.}}
Finally, we apply the QI tools from the Krylov framework to the dipole cascade model both with and without recombinations. More precisely, we evaluate the K-complexity \eqref{eq:krylovC}, the K-entropy \eqref{eq:KEntropy} and the K-variance \eqref{KVariance}. Moreover, we evaluate the capacity of entanglement \eqref{eq:KCapacity}, which contains information about the width of the eigenvalue distribution of the reduced density matrix \cite{Yao_2010} and the purity \eqref{eq:Purity}, which is strictly 1 for pure states and deviates from this value as the system becomes mixed.
\begin{figure}[t]
    \centering
    \includegraphics[width=0.49\textwidth]{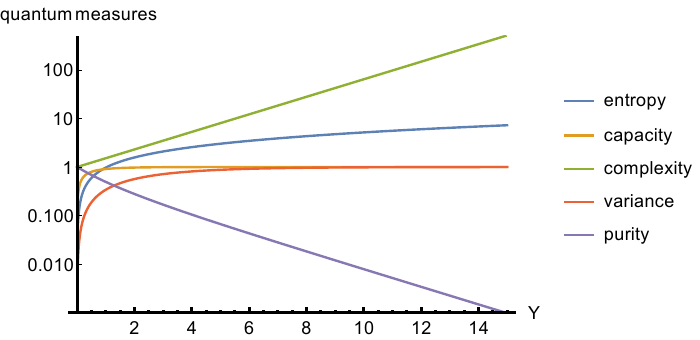}
    \caption{Quantum measures applied to solutions of eq. (\eqref{eq:cascade1}). The plot is for $\alpha_s=0.15$.}
    \label{fig:results1}
\end{figure}
\begin{figure}[t]
    \centering
    \includegraphics[width=0.49\textwidth]{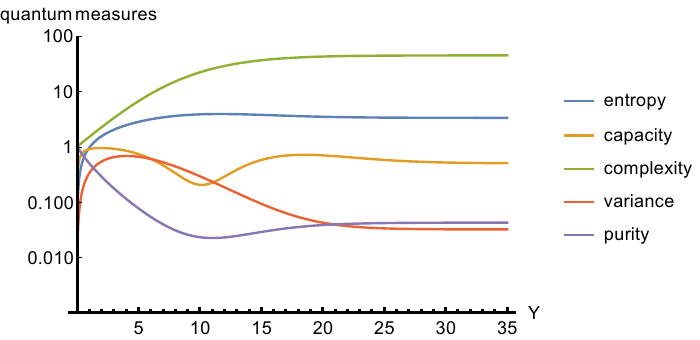}
    \caption{Quantum measures applied to solutions of eq. (\eqref{eq:cascade2}). Plot for $\alpha_s=0.15$.}
    \label{fig:qmeasures}
\end{figure}\\
First, in fig.\,\ref{fig:results1}, we plot the QI measures that characterize the dipole cascade eq. \eqref{eq:cascade1} using the exact  solution of \cite{Mueller:1994gb}. We confirm previous results that the entropy grows linearly with Y and reaches the maximally entangled state \cite{Kharzeev:2017qzs} at large rapidities. This is also supported by the value of purity tending to zero.
The rate of the growth of entropy is given by $\lambda$ and the complexity grows exponentially. 
We also find that the capacity of entanglement measuring entanglement spreading saturates at unity. \\
Next, we plot the same measures for the solutions of the dipole model that accounts for recombinations.  The complexity grows initially, but saturates at large rapidities \footnote{Note that the enormous rapidity range that we use is relevant to our theoretical study of the models. The phenomenologically relevant rapidity range is  $Y < 7$. }. From the QCD perspective, this behavior confirms an expected feature of the interplay between the splitting and recombination terms in the eq. \eqref{eq:cascade2}.
The entropy, however, develops a maximum as  observed in \cite{Hagiwara:2017uaz} and saturates at larger rapidities. We note that such behavior is consistent with gluon densities obtained with QCD evolution equations that take into account transverse degrees of freedom \cite{Hentschinski:2022rsa}. This is also consistent with the evolution of the capacity of entanglement, which has the minimum (see Appendix C for a plot in the linear scale). Interestingly, the system does not reach a state where the purity drops to zero, indicating a departure from the maximally entangled state. 
Overall, we interpret this behavior as an indication of ``classicalization" \cite{Dvali:2021ooc}, i.e. the emergence of scale in the system. In our case, this is the value of rapidity for which the capacity has a minimum and the complexity saturates.
This is also consistent with the results shown in fig. \eqref{fig:qmeasures}, where the negativity of the system with saturation is lower.
\paragraph*{\textbf{Conclusions and discussion.}}
In conclusion, we have established a connection between the framework of the evolution of states with dynamical $SL(2,R)$ symmetry in the Krylov basis and the evolution of dipoles in the 1+1 dimensional Mueller dipole cascade model. 
This new connection allowed us to relate the K-complexity to the number of dipoles and the K-entropy to the entanglement entropy of dipoles in DIS discussed in \cite{Kharzeev:2017qzs}. 
The results with the simple dipole model in 1+1 D  have a chance to hold more generally. In particular, the dipole model 
in 3+1 D has formally the same structure as the 1+1 D model. This suggests that Krylov space methods should also be applicable and that the relation between quantum complexity and the gluon density can be generalized more broadly. Indeed, conformal symmetry and its consequences in QCD has been discussed before \cite{Braun:2003rp} and it will be very important to understand how Krylov methods fit into this story\footnote{We thank the anonymous referee for point this to us.}.\\
We have also imported some of the new, useful QI tools to explore more fine-grained properties of the dipole cascade models. The first of the models features linear growth of the entropy with rapidity \cite{Kharzeev:2017qzs,Gursoy:2023hge,Kutak:2023cwg} and exponential growth of complexity, while the second model shows the saturation of these measures. Interestingly, from the perspective of quantum complexity, we usually think about the saturation of complexity as a consequence of the finite Hilbert space. However, in the QCD model, the saturation comes from the interplay between splitting and recombination of partons (both infinite dimensional). It will be very interesting to understand this mechanism from a more QI perspective, and we leave it as an important future problem.

 We also analyzed the variance, capacity and purity measures. The models with and without saturation predict very different behavior of these new, potentially measurable quantities. Note, for example, that the evolution of capacity shows a characteristic minimum at the maximum value of entropy. Therefore, it can potentially be a more sensitive probe of the degree of parton entanglement to be detected. 

Last but not the least, we are certain that these QI tools will not only be useful for future analysis of the entanglement structure governed by more realistic QCD evolution equations but may also be measurable under the same assumptions as the hadronic entropy. 
\paragraph*{Acknowledgements.}
We are grateful to Vijay Balasubramanian, Martin Hentschinski, Javier Magan, Dimitrios Patramanis and Douglas Stanford for discussions and comments. We thank the anonymous referee for interesting perspective on our work and outline of generalising it to more interesting operators in QCD. PC is supported by NAWA “Polish Returns 2019” PPN/PPO/2019/1/00010/U/0001 and NCN Sonata Bis 9 2019/34/E/ST2/00123 grants. PC would also like to thank the Isaac Newton Institute for Mathematical Sciences, Cambridge, for their support and hospitality during the ``Black Holes: bridges between number theory and holographic quantum information" program, where work on this paper was undertaken and presented as well as KITP, UCSB for support during the final stages of this project. KK acknowledges
the European Union’s Horizon 2020 research and innovation program under grant agreement No. 824093.
%
\appendix
\section{General SL(2,R) solutions} 
In this appendix, we provide a brief summary of the analytical solution \eqref{SolSL2R} and use it to evaluate the QI tools discussed in the main text.\\
A general class of analytical solutions to equation \eqref{SchrodingerEq} can be obtained when 
\be
a_n=\gamma(n+h),\qquad b_n=\alpha\sqrt{n(2h+n-1)},
\ee
for parameters $\gamma$ and $\alpha$ that depend on the physical details of the initial state and evolving Hamiltonian. The parameter $h$ corresponds to the conformal dimension of the operator the growth of which we are considering. As described in \cite{Caputa:2021sib}, these Lanczos coefficients reflect the underlying dynamical SL(2,R) symmetry of the Krylov basis for a given physical problem.\\
In the following, we focus on the case when $\gamma=0$, for which the solution $\psi_n(t)$ yields the probabilities
\be
 p_n(t)=|\psi_n(t)|^2=\frac{\Gamma(2h+n)}{n!\Gamma(2h)}\frac{\tanh^{2n}(\alpha t)}{\cosh^{4h}(\alpha t)}.
\ee
It is useful to think of this example as a state
\be
\ket{\psi(t)}=e^{i\alpha (a^\dagger b^\dagger+ab)t}\ket{0}\otimes\ket{0},\label{CS2}
\ee
where the operators in the exponent $L_{-1}=a^\dagger b^\dagger$ and $L_1=ab$, are the raising and lowering operators of the SL(2,R) Lie algebra. The third generator $L_0=\frac{1}{2}(a^\dagger a+b^\dagger b+1)$ and $\ket{0}\otimes\ket{0}$ is its eigenstate $L_0\ket{h}=h\ket{h}$ with $h=1/2$\footnote{General $h$ can be obtained by acting with the same exponent \eqref{CS2} on the initial state with $k=2h-1$ difference between the modes. These are the k-added or subtracted states in the jargon of quantum optics.}. This is simply a special case of the coherent state of the SU(1,1) (SL(2,R)) algebra and we can expand it as
\be
\ket{\psi(t)}=\sum^\infty_{n=0}\frac{\tanh^n(\alpha t)}{\cosh^{2h}(\alpha t)}\sqrt{\frac{\Gamma(2h+n)}{n!\Gamma(2h)}}\ket{n}\otimes\ket{n}.
\ee
We think of this state as a Schmidt decomposition in the product of two Hilbert spaces, so that tracing out over one of them yields the density matrix
\be
\rho(t)=\sum^\infty_{n=0}p_n(t)\ket{n}\bra{n}.
\ee
It is useful to rewrite these probabilities in terms of the rapidity variable $Y$ 
\be
p_n(Y)=\frac{\Gamma(2h+n)}{n!\Gamma(2h)}(e^{-\alpha Y})^{2h}\left(1-e^{-\alpha Y}\right)^n.
\ee
This probability satisfies a cascade equation
\be
\partial_Y p_n(Y)=\alpha (n+2h-1) p_{n-1}(Y)-\alpha(n+2h)p_{n}(Y),
\ee
and for $h=1/2$ we recover our example \eqref{EqProbY}.\\
One way to understand this equation is simply by analogy with the Schrodinger equation for the wave function and the continuity equation for the probabilities. In fact, we can simply write our Schrodinger equations for the Krylov wave function and for its complex conjugate, multiply them by the conjugate wave functions and add them. The left hand side is the derivative of the probability $p_n(t)$ and the right side is a combination of different wave functions and their conjugates. We can check that our SL(2,R) solution has the property that the right hand side can be rewritten as the above equation times overall function of $t$. Finally, changing the variables to $Y$ removes this factor and yields the above equation.\\ 
Next, to present the analytical results for the QI measures, it is  useful to parameterize our probability distribution by the variable
\be
z\bar{z}=1-e^{-\alpha Y}=\tanh^2(\alpha t),\label{Varzzb}
\ee
such that
\be
p_n=(1-z\bar{z})(z\bar{z})^{n}.
\ee
This way, we can rewrite Krylov complexity as
\be
C_K=\langle n\rangle=\frac{z\bar{z}}{1-z\bar{z}},
\ee
while the K-entropy is given by
\be
S_K=\log \left(\frac{z\bar{z}}{1-z\bar{z}}\right)-\frac{\log(z\bar{z})}{1-z\bar{z}}.
\ee
Clearly, we have the relation
\be
S_K=\log \langle n\rangle-\frac{\log(z\bar{z})}{1-z\bar{z}},
\ee
which for late times, as $z\bar{z}\to 1$, relates $S_K$ and K-complexity $\langle n\rangle$. The variance is simply given by 
\be
\delta_K^2=(z\bar{z})^{-1}.
\ee
Moreover, the Renyi entropies 
\be
S^{(m)}=\frac{1}{1-m}\log\left(\frac{(1-z\bar{z})^m}{1-(z\bar{z})^m}\right),
\ee
give rise to the capacity of entanglement
\be
C_E=\frac{z\bar{z}\log(z\bar{z})^2}{(1-z\bar{z})^2}.
\ee
Finally, we can also evaluate the purity as
\be
\gamma_K=\frac{1-z\bar{z}}{1+z\bar{z}}.
\ee
Plugging $z\bar{z}$ from \eqref{Varzzb} reproduces all our results from the main text.
\section{The continuum limit}
It is often useful to consider the continuum limit of the discrete chain equation \cite{Barbon:2019wsy} and we repeat this procedure for the general model with saturation. Namely, starting from \eqref{eq:cascade2}, we can write it as
\be
 \partial_Y p_n= b_{n-1}p_{n-1}-b_n p_n+c_{n+1}p_{n+1} - c_{n}p_n,
\ee
where in our explicit equation we have $b_n=\lambda n$ and $c_n=\beta n(n-1)$, but we keep them general for the moment.\\
Next, we define the continuous variable $x\equiv\epsilon\, n$, so that
\be
p_n(Y)\equiv p(x,Y),\qquad p_{n\pm1}=p(x\pm \epsilon,Y).
\ee
Moreover, we write
\be
\epsilon b_n\equiv v(x),\qquad \epsilon c_n \equiv w(x),
\ee
as well as
\be
\epsilon b_{n- 1}=v(x-\epsilon),\qquad \epsilon c_{n+ 1}=w(x+\epsilon).
\ee
After expanding for $\epsilon\simeq 0$, we derive the continuum equation
\be
\partial_Y p(x,Y)+\partial_x j(x,Y)=0,
\ee
where
\be
j(x,Y)=\left(w(x)-v(x)\right)p(x,Y).
\ee
For $\beta=0$ and $v(x)=\lambda x$, we reproduce \cite{Barbon:2019wsy}, but the generalized equation is responsible for the effective $\tilde{v}(x)=w(x)-v(x)$. For example, we can take the initial Gaussian profile
\be
p(x,0)=\sqrt{\frac{2}{\pi}}\frac{1}{\sigma}e^{-\frac{x^2}{2\sigma^2}},
\ee
so that, if $v(x)=\lambda x$ and $\beta=0$, the solution becomes
\be
p(x,Y)=\sqrt{\frac{2}{\pi}}\frac{1}{\sigma}\exp\left(-\lambda Y-e^{-2\lambda Y}\frac{x^2}{2\sigma^2}\right).
\ee
This gives the exponential growth of complexity
\be
\langle x\rangle=\int^\infty_0 x\, p(x,Y)dx=\sqrt{\frac{2}{\pi}}\sigma e^{\lambda Y}.
\ee
On the other hand, if $w(x)=v(x)$, we get a stationary solution $p(x,Y)=p(x,0)$, which leads to constant
\be
\langle x\rangle=\sqrt{\frac{2}{\pi}}\sigma.
\ee
Recall that the discrete stationary solution of the general equation \eqref{eq:cascade2} is given by \cite{Bondarenko:2006rh}
\be
p_n(Y)=\frac{N^n}{n!}e^{-N},\qquad \langle n\rangle=N=1/\alpha^2_s,
\ee
and corresponds to the continuum intuitions above. 
\section{Additional plots}
\begin{figure}[b!]
    \centering  \includegraphics[width=0.49\textwidth]{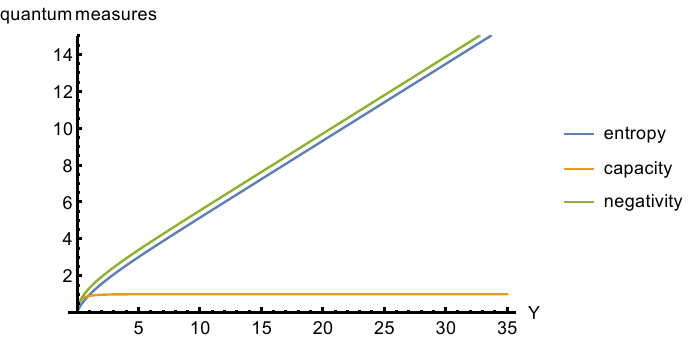}
\includegraphics[width=0.49\textwidth]{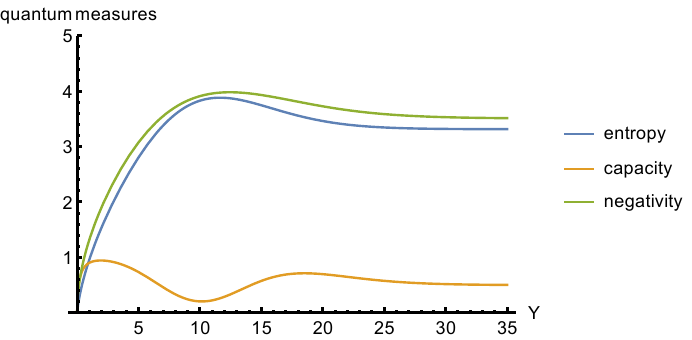}
    \caption{Plot of entropy, capacity of entanglement and negativity extracted from solutions of eqs. \eqref{eq:cascade1} and \eqref{eq:cascade2}.}.
    \label{fig:qmeasures}
\end{figure}
In this appendix, we provide additional figures in linear scale that complement those in the main body of the paper. In particular, for the two QCD models of the main text,  together with the entanglement entropy and the capacity of entanglement, we also evaluate the logarithmic negativity defined as \cite{Vidal:2002zz,Plenio:2005cwa}
\begin{equation}
    {\cal N}(Y)=2\ln\left(\sum_n\sqrt{p_n(Y)}\right).
\end{equation}
Clearly, the maximum and decrease of the entanglement entropy is now also visible in the evolution of the capacity of entanglement.
The logarithmic negativity is positive, which means that the entropy is entanglement entropy, i.e. quantum entropy. 

\end{document}